# Self-adaptive single frequency laser assisted by distributed feedbacks

Fuhui Li[1,2], Ligang Huang[1,2], Laiyang Dang[1,2], Guolu Yin[1,2], Lei Gao[1], Tianyi Lan[1], Yujia Li[1], Lidan Jiang[1], Leilei Shi[1], and Tao Zhu[1,*]

[1]Key Laboratory of Optoelectronic Technology & Systems, Education Ministry of China, Chongqing University, Chongqing 400044, China

[2]These authors contributed equally

* E-mail: zhutao@cqu.edu.cn

High-coherence light sources with an ultra-narrow linewidth are of considerable research interests in numerous fields, such as laser interferometer gravitational-wave observatory, laser radio, etc. Herein, commencing from the in-depth analysis for consumption and replenishment law of the carriers in the gain-band under excitation of the feedback signal, a compression idea to compress extremely laser linewidth assisted by distributed feedback has been proposed. Consequentially, a novel laser configuration is demonstrated. A cavity mode signal matching with the output wavelength is fed back into the main cavity by the distributed feedback structure to provide an excitation signal required for laser gain. Moreover, the corresponding experimental investigation is conducted based on an on-chip laser system with distributed feedback. Eventually, an ultra-narrow linewidth laser with a side mode suppuration ratio of 80 dB, a linewidth of 10 Hz, a Lorentzian linewidth of sub Hz, and a relative intensity noise (RIN) of -150 dB/Hz is obtained under normal conditions. The proposed idea is valid in any other gain-types lasers with diverse wavelengths, which also provides a new perspective for other laser parameters for extreme modulation.

With the rapid development of optical information technology, high-coherence light source has become the kernel carrier to promote the development of optical communication [1-3], optical precision measurement [4-7], and signal synthesis [8-10] due to its ultra-long coherent length and extremely low phase noise. Moreover, spectrum linewidth acting as one of the important parameters employed to characterize the coherence of light source has caused considerable attention in the field of academic research and industry [11-13]. In order to obtain the high-coherence light sources, many laser configurations have been proposed, which can be attached to two main methods, i.e., external circuit control method [14-16] and optical feedback method [17-19]. For the external circuit control method, the improvement of coherence mainly depends on the laser frequency stabilization technology based on electrical feedback, which is restricted by the external control element and the operating environment of the light sources. The optical feedback method is an optical power feedback based on a fixed external cavity, which mainly depends on the increase of cavity length to improve the coherence of the laser source. However, the increase of cavity length will be vulnerable to environmental perturbations [20], thermal dynamics noise [21], and other affects, which would restrict the further improvement of the laser coherence. In addition, although the high-$Q$ microcavity acting as an external feedback device has also been investigated widely, the potential applications are limited by its selectivity of wavelength to the main cavity [22, 23]. Therefore, it remains a challenge that how to further extremely compress the laser linewidth to obtain an ultra-high coherence light source.

In view of the difficulties brought by the traditional narrow linewidth laser, a compression idea employed to realize the extreme compression of laser linewidth is proposed independently by deeply analyzing the consumption and replenishment law of the carriers in the gain band under the excitation of feedback signal, which can be declared as a novel laser structure based on the distributed feedback structure. In this kind of laser cavity structure, a cavity mode signal matched with output wavelength is fed back into the main cavity by the distributed feedback structure to provide an excitation signal required for laser gain, which would open up a new degree of freedom for realizing the extreme compression of laser linewidth. On this basis, an experimental investigation employed to verify the proposed compression idea is conducted by utilizing an on-chip laser system with distributed feedback. Eventually, an ultra-narrow linewidth laser with a side mode suppression ratio (SMSR) greater than 80 dB, output linewidth of 10 Hz, and a relative intensity noise (RIN) less than -150 dBm/Hz is successfully obtained under normal conditions. The compression idea based on the distributed feedback in this work will open up a new perspective for improving and obtaining other types of high coherence laser sources, which is also of a great reference significance for other laser parameters to realize the extreme modulation. To explore the physical factors that restrict the extreme compression of laser linewidth, an in-depth analysis on the suppression principle of spontaneous emission has been conducted by commencing from the consumption and replenishment law of the carriers in the gain-band under excitation of the feedback signal. Herein, the effective suppression of spontaneous emission generated in the gain process by utilizing the excitation signal from the external feedback is a core of realizing the extreme compression of laser linewidth. Subsequently, a compression idea employed to realize the extreme compression for laser linewidth by a distributed feedback is propose independently, and a novel laser configuration is demonstrated, as shown in Fig.1, which is composed of a main laser cavity and an external cavity with distributed feedback characteristic. A broad-band gain can be provided by the gain medium under diverse pumping techniques and an initial lasing signal would output from one side of the main cavity after countless optical oscillation and longitudinal mode competition. The output signal will be injected into the distributed feedback structure through a coupling system. According to the distributed feedback characteristics of the external cavity, a cavity mode signal matched with the output wavelength is generated and fed back into the main cavity to provide an excitation signal required for laser gain.

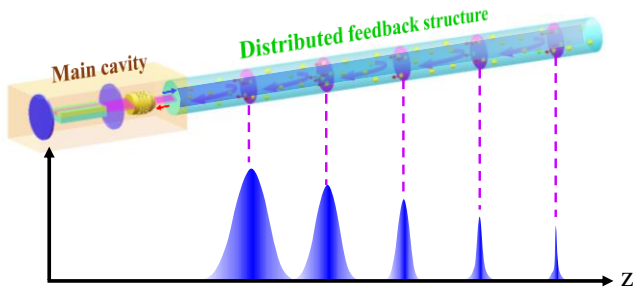

**Fig.1** A novel laser structure assisted by a distributed feedback.

Based on this laser structure, apart from the matching characteristics for main cavity wavelength, the spectral linewidth of the feedback-mode signal decreases with the increase of the distance of the feedback plane, which can be declared as followings:

$$\Delta v_D(z) = \frac{c\left[1-\beta\Theta S\exp(-\alpha z)\right]}{2\pi z} \quad (1)$$

where $c$ denotes light speed, $\beta$ denotes the feedback coefficient of each particle, $\Theta$ denotes the particle concentration on the feedback plane, $S$ is effective area of each feedback plane, $\alpha$ is the transmission loss of photons in feedback structure. In this process, the cavity mode signal with a purer spectrum during each oscillation is used as an excitation signal in the laser gain process to suppress spontaneous emission as much as possible, which means the laser linewidth will be extremely compressed under the excitation of cavity mode signal matching the main cavity. In order to reveal the characteristics of spontaneous emission excited by the cavity mode signal, the spontaneous emission rate has been deeply analyzed, which can be described as followings:

$$R_{21}(v) = n_2 A_{21}(v)$$
$$= \frac{c\chi N_m(v)}{n_{eff}l} - N_B(v) - \Delta n W_{21}(v) \quad (2)$$

where $\chi$ donates the total coefficient loss, $N_m(v)$ denotes total photon number density in the main cavity, $n_{eff}$ denotes effective refractive index, $N_B(v)$ denotes photon number corresponding to the feedback-mode signal from the distributed feedback structure, $\Delta n$ is the number of reversed-carriers in the energy band, $n_2$ donates the carrier number density at lasing energy-level, $W_{21}$ donates the stimulated radiation probability corresponding to the feedback-mode signal. In the process of laser oscillation based on this laser structure, due to the initial laser signal from the main cavity is a wide spectrum signal with continuous distribution in frequency domain, the output linewidth after improvement by the excited signal provided by the distributed feedback structure can be expressed as follows:

$$\Delta v = \frac{c(1+\eta^2)}{4\pi n_{eff}}\left[\frac{\chi}{l} - g - \frac{T\beta\Theta S}{2l\alpha}\left[1-\exp(-2\alpha L)\right]\right] \quad (3)$$

where the $\eta$ denotes the linewidth enhancement factor, $g$ denotes the gain factor, $T$ denotes the a transmissivity corresponding to the output end of the main cavity, and $L$ is the length of the feedback structure.

Based on **Eq**. (3), the corresponding simulation results is shown in Fig. 2. As shown in Fig.2 (a), the compression law of laser linewidth with the length of distributed feedback is demonstrated, where the spectrum linewidth is marked by the brightness of curve color. It is obvious that the output spectrum becomes narrower and narrower with the increase of the length. According to the change trend of the curve, there is a sharp compression for the output spectrum in the range of 30 m. Subsequently, the linewidth gradually tends to be an ideal state at a slow rate. It suggests that the cavity-mode signal from the distributed feedback structure not only plays the role of the optical power feedback, also can realize a depth-compression for laser linewidth by suppressing the spontaneous emission during the laser oscillation. Figure 2(b) shows the variation characteristics of laser linewidth at different feedback ratios. Similarly, the output spectrum is marked by the brightness of curve color. It can be see that the spectrum is narrower and narrower with the increase of the feedback ratio, which suggests that the size of distributed feedback structure can be reduced by enhancing the feedback ratio. Compared with the traditional optical feedback based on the fixed cavity, the proposed distributed feedback structure can realizing the extreme compression of laser linewidth by provide a cavity mode signal matched with the main cavity wavelength to suppress the spontaneous emission.

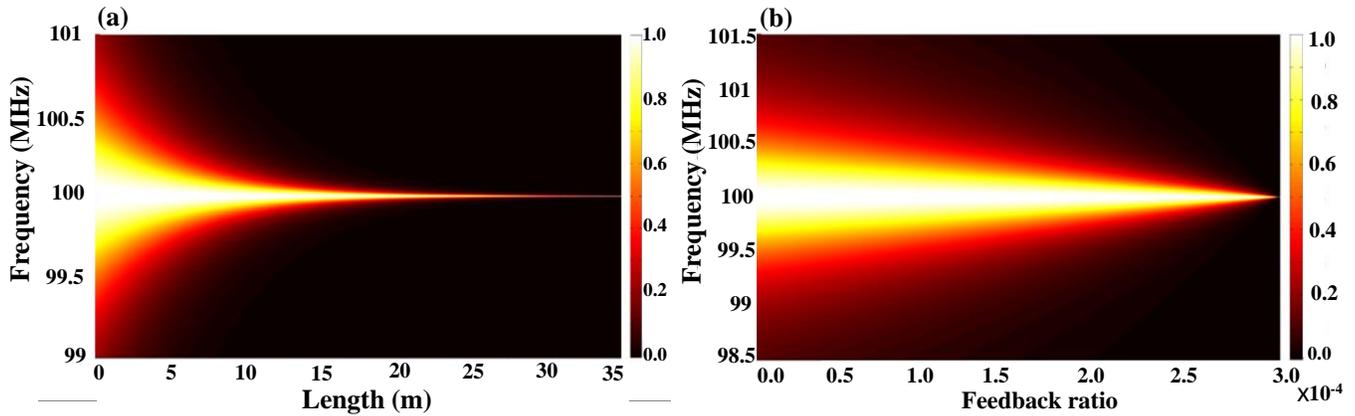

**Fig.2** Simulation results of linewidth evolution. (a) Compression characteristics of laser linewidth caused by the length change of distributed feedback, (b) Compression characteristics of laser linewidth caused by the feedback ratio of distributed feedback

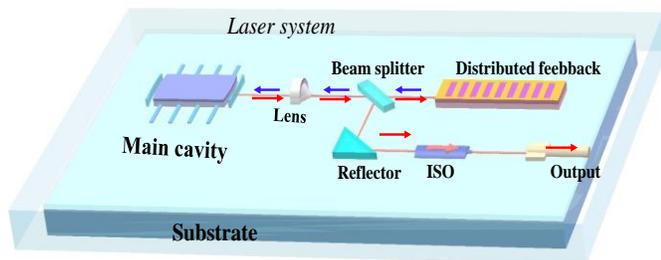

**Fig.3** An on-chip laser system based on distributed feedback external cavity structure

In order to verify the proposed idea employed to realize the extreme compression of laser linewidth, an on-chip laser system based on distributed feedback external cavity structure is proposed, as shown in Fig. 3, which is mainly composed of a main laser cavity and an external cavity with distributed feedback. The laser main cavity is driven by current to provide an initial optical signal with an output linewidth of MHz magnitude. The signal output from the main cavity is injected to the distributed feedback structure via a beam splitter. And then, a cavity mode signal matched with the lasing wavelength is generated and fed back into the main cavity to provide an excitation signal required by laser gain. The other port of the beam splitter is used to the laser output. An optical isolator is introduced to prevent the influence from the reflection of other parts of the experimental system on the measurement results. Due to the suppression of spontaneous emission in laser gain process by feedback signal from the distributed feedback structure, an ultra-narrow linewidth laser will obtained eventually. In addition, to ensure the accuracy of characterizing the output laser, frequency noise measurement method based on differential phase demodulation and beat frequency method by using two lasers with the same parameters are introduced.

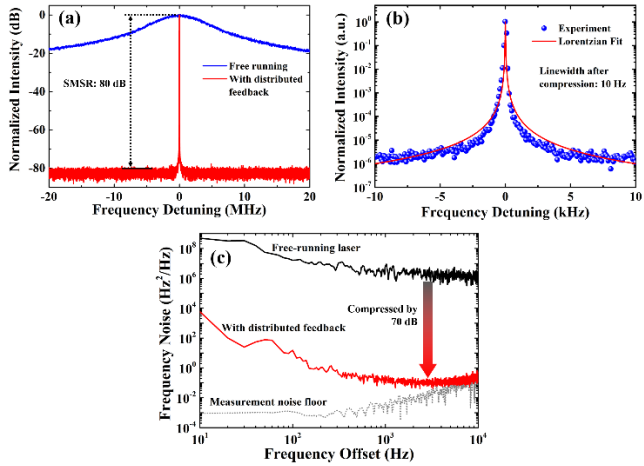

**Fig.4** (a) Comparison curves of the frequency spectrum from beat frequency signal, (b) Lorentz fitting curve of the linewidth with a distributed feedback, (c) Comparison curves of the frequency noise PSD.

Based on the proposed laser structure, the experimental results are shown in Fig. 4. To directly demonstrate the ultra-narrow property of the laser assisted by the distributed feedback, two identical lasers have been built. The beat-frequency signal between two lasers are measured by a real-time oscilloscope with a bandwidth of 20 GHz. The frequency spectrum of the beat-frequency signal is shown in Fig. 4(a), where the electric spectrum corresponding to the free-running DFB laser and the laser assisted by a feedback power ratio of -45 dB are respectively reflected by blue and red curves. It is obvious that the spectral linewidth of red curve is much smaller than that of blue curve. In addition, a side mode suppression ratio (SMSR) larger than 80 dB is demonstrated from the curve in a span of 40 MHz. The fitting result of the output linewidth corresponding to the red curve in Fig. 4(a) is approximately 10 Hz, as shown in Fig. 4(b). Such a linewidth value has already reached the top level in the state of art under the room-temperature condition. Compared with the free-running DFB, the linewidth is successfully compressed by 5 orders. The frequency noise power spectral density (PSD) spectrum is then obtained with the demodulated transient frequency by fast Fourier transformation (FFT), which is shown in Fig. 4(c). Herein, the frequency noise PSD corresponding to the free-DFB laser and the laser assisted by distributed feedback are marked by black and red curves, respectively. Obviously, the white noise floor marked by red curve is reduced by 70 dB compared with that of the black curve, and it is approaching sub-$Hz^2$/Hz magnitude, which suggests that the output linewidth is about several Hz. In addition, the gray curve denotes the measurement noise floor, it can be see that the white noise floor revealed by red curve and gray curve almost coincide, which means the linewidth is approximately approaching the measurement limit.

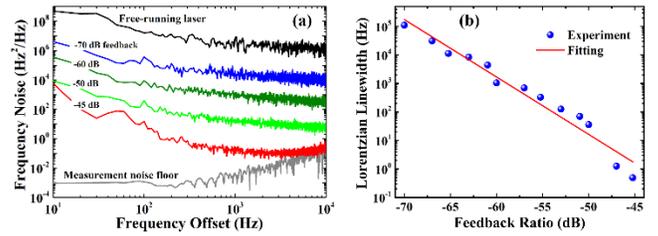

**Fig.5** (a) Evolution curves of the frequency noise PSD under different feedback ratio, (b) Evolution curves of output linewidth with the feedback ratio.

To further reveal the influence of feedback ratio from the distributed feedback structure on laser output characteristics, the corresponding experimental results are shown in Fig. 5. The frequency noise PSD curves measured under the different feedback ratio are shown in Fig. 5(a). Obviously, the frequency noise floor within the white noise area around 10 kHz gradually varies from $10^6$ Hz order to sub-Hz order, which means the intrinsic linewidth of the laser determines the Lorentzian shape is decreasing with the increasing of the feedback ratio. The corresponding Lorentzian linewidth variation curve is shown in Fig. 5(b). The change tendency is also coincided with the previous theory expectation. That is to say, with a stronger distributed feedback, the inherent laser frequency noise can be well restricted.

Figure 6 shows the dynamic process of self-adaptive compression. From Fig. 6(a), we can see that the transient spectrum has been extremely compressed with the increase of the time when switching on the feedback by an electro-optic modulator (EOM), especially in the range marked by yellow curve. The corresponding Lorentzian linewidth is demonstrated in Fig. 6(b). Obviously, the linewidth is sharply compressed from MHz order to 10-Hz order in a time range of 1.1 ms, which approaches the coherence time of tens-of-Hz narrow-linewidth lasers. In addition, to further reveal the dynamic process of linewidth compression in tuning process for main cavity wavelength, the Fig. 6(c) is introduced to demonstrate the transient spectrum, and the corresponding Lorentzian linewidth when tuning the frequency of the main laser cavity is shown in Fig. 6(d). The frequency switching time is $\Delta t$=2.5 ms, which contains three processes, i.e., decompression, free running and recompression. Obviously, the linewidth increases sharply when the feedback cavity signal is detuning for the main laser, which means that the feedback signal matched with the main cavity wavelength plays a key role in extreme compression of laser linewidth.

In order to further demonstrate the stability of the laser with distributed feedback, the relative intensity noise (RIN) is also measured, as shown in Fig. 7. The red curve and the blue curve respectively represent the RIN spectrum with and without distributed feedback. It can be see that the RIN spectrum marked by the red curve is obviously improved by 20 dB. It suggests that the influence from external disturbance can be also suppressed by the feedback signal matched with the main cavity wavelength in this laser structure.

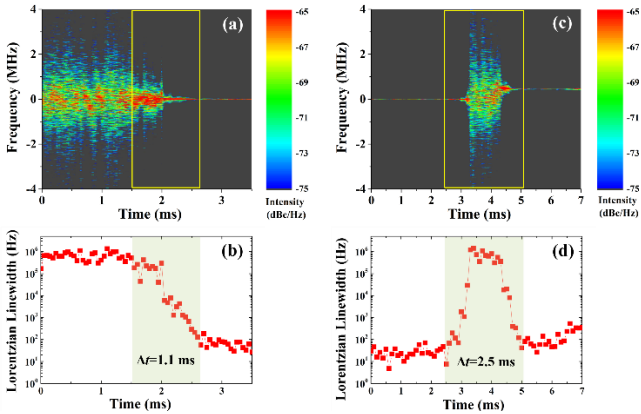

**Fig.6** Self-adaptive compression process of laser linewidth. (a) The transient spectrum and (b) the corresponding Lorentzian linewidth when switching the feedback on. (c) The transient spectrum and (d) the corresponding Lorentzian linewidth when tuning the frequency of the main laser cavity.

Although the proposed compression idea based on a distributed feedback opens up a new perspective to obtain ultra-high coherent light sources for other gain wavelength or to realize the extreme control of other laser parameters, it is very desirable to further reveal and verify the dynamic characteristics of carrier transition in gain band by means of utilizing microscopic measurement. Moreover, although the corresponding experimental investigation has been conducted by utilizing an on-chip laser system with distributed feedback, the current measurement results is not a perfect representation for this proposed idea. Provided an appropriate distributed feedback mechanism could be sought or machined, it will be of great significance to obtain the compact and integrated high coherent light source.

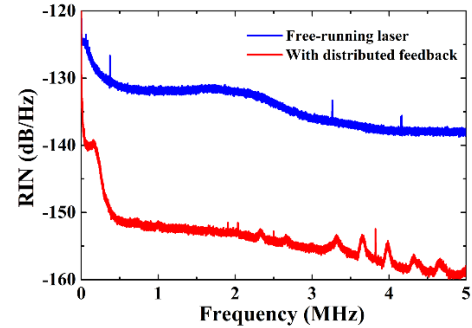

**Fig.7** Comparative curves of the RIN spectrum

In conclusion, we have independently proposed a compression idea employed to extremely compress laser linewidth by deeply analyzing the consumption and replenishment law of the carriers in the gain-band under the excitation of feedback signal. Consequentially, we further demonstrate a novel configuration of ultra-narrow linewidth laser, which is composed of a main laser cavity and an external cavity with a distributed feedback. The distributed feedback structure is employed to feedback a cavity mode signal matched with output wavelength to provide an excitation signal required for laser gain, resulting in the coherence of laser could be extremely enhanced. Moreover, an experimental investigation employed to verify the proposed compression ideal is conducted by utilizing an on-chip laser system with distributed feedback. Eventually, an ultra-narrow linewidth laser with a SMSR greater than 80 dB, output linewidth of 10 Hz, and a RIN less than -150 dB/ Hz is successfully obtained under normal conditions. The proposed idea and the laser configuration in this work would open up a new perspective for improving and obtaining other types of high coherence laser sources, which is also of a great reference significance for other laser parameters to realize the extreme modulation.